\documentclass{elsart3p}

 \usepackage{graphicx}

\usepackage{amssymb}

\begin{document}

\begin{frontmatter}

% Title, authors and addresses
\date{15 December 2007}

\title{The Fast Read-out System for the MAPMTs of COMPASS RICH-1}

\author[saclay]{P.Abbon},
\author[alessandria]{M.Alexeev\thanksref{leave-jinr}},
\author[munich]{H.Angerer},
\author[infn-trieste]{R.Birsa},
\author[lip]{P.Bordalo\thanksref{also-ist}},
\author[trieste]{F.Bradamante},
\author[trieste]{A.Bressan},
\author[torino]{M.Chiosso},
\author[trieste]{P.Ciliberti},
\author[infn-torino]{M.L.Colantoni},
\author[saclay]{T.Dafni},
\author[infn-trieste]{S.Dalla Torre},
\author[saclay]{E.Delagnes},
\author[infn-torino]{O.Denisov},
\author[saclay]{H.Deschamps},
\author[infn-trieste]{V.Diaz},
\author[torino]{N.Dibiase},
\author[trieste]{V.Duic},
\author[erlangen]{W.Eyrich}, 
\author[torino]{A.Ferrero},
\author[prague]{M.Finger},
\author[prague]{M.Finger Jr},
\author[freiburg]{H.Fischer},
\author[munich]{S.Gerassimov},
\author[trieste]{M.Giorgi},
\author[infn-trieste]{B.Gobbo},
\author[freiburg]{R.Hagemann},
\author[mainz]{D.von~Harrach},
\author[freiburg]{F.H.Heinsius},
\author[bonn]{R. Joosten},
\author[munich]{B.Ketzer},
\author[cern]{V.N. Kolosov\thanksref{leave-ihep}},
\author[freiburg]{K.K\"onigsmann}, 
\author[munich]{I.Konorov},
\author[liberec]{D.Kramer},
\author[saclay]{F.Kunne},
\author[erlangen]{A.Lehmann},
\author[trieste]{S.Levorato},
\author[infn-torino]{A.Maggiora},
\author[saclay]{A.Magnon},
\author[munich]{A.Mann},
\author[trieste]{A.Martin},
\author[infn-trieste]{G.Menon},
\author[freiburg]{A.Mutter},
\author[bonn]{O.N\"ahle},
\author[freiburg]{F.Nerling}, 
\author[saclay]{D.Neyret},
\author[alessandria]{D.Panzieri},
\author[munich]{S.Paul},
\author[trieste]{G.Pesaro},
\author[erlangen]{C.Pizzolotto},
\author[liberec,infn-trieste]{J.Polak},
\author[saclay]{P.Rebourgeard},
\author[saclay]{F.Robinet},
\author[torino]{E.Rocco},
\author[trieste]{P.Schiavon}, 
\author[freiburg]{C.Schill \corauthref{Schill}},
\author[erlangen]{P.Schoenmeier},
\author[erlangen]{W.Schr\"oder},
\author[lip]{L.Silva},
\author[prague]{M.Slunecka},
\author[trieste]{F.Sozzi},
\author[prague]{L.Steiger},
\author[liberec]{M.Sulc},
\author[liberec]{M.Svec},
\author[trieste]{S.Takekawa},
\author[infn-trieste]{F.Tessarotto},
\author[erlangen]{A.Teufel},
\author[freiburg]{H.Wollny}
\address[alessandria]{INFN, Sezione di Torino and University of East Piemonte, Alessandria, Italy}
\address[bonn]{Universit\"at Bonn, Helmholtz-Institut f\"ur Strahlen- und Kernphysik, Bonn, Germany}
\address[cern]{CERN, European Organization for Nuclear Research, Geneva, Switzerland}
\address[erlangen]{Universit\"at Erlangen-N\"urnberg, Physikalisches Institut, Erlangen, Germany}
\address[freiburg]{Universit\"at Freiburg, Physikalisches Institut, Freiburg, Germany}
\address[liberec]{Technical University of Liberec, Liberec, Czech Republic}
\address[lip]{LIP, Lisbon, Portugal}
\address[mainz]{Universit\"at Mainz, Institut f\"ur Kernphysik, Mainz, Germany}
\address[munich]{Technische Universit\"at M\"unchen, Physik Department, Garching, Germany}
\address[prague]{Charles University, Praga, Czech Republic  and JINR, Dubna, Russia}
\address[saclay]{CEA Saclay, DSM/DAPNIA, Gif-sur-Yvette, France}
\address[torino]{INFN, Sezione di Torino and University of Torino, Torino, Italy}
\address[infn-torino]{INFN, Sezione di Torino, Torino, Italy}
\address[trieste]{INFN, Sezione di Trieste and University of Trieste, Trieste, Italy}
\address[infn-trieste]{INFN, Sezione di Trieste, Trieste, Italy}

\vspace{-0.8cm}

\thanks[leave-jinr]{on leave  from JINR, Dubna, Russia}
\thanks[also-ist]{also at IST, Universidade T\'ecnica de Lisboa, Lisbon,
\newline \hspace*{0.26cm}\mbox{Portugal}}
\thanks[leave-ihep]{on leave  from IHEP, Protvino, Russia}

\corauth[Schill] {Corresponding author, email:
\texttt{Christian.Schill@cern.ch}\\}

\begin{abstract}
A fast readout system for the upgrade of the COMPASS RICH detector has
been developed and successfully used for data taking in 2006 and 2007. The new
readout system for the multi-anode PMTs in the central part of the photon
detector of the RICH is based on the high-sensitivity MAD4
preamplifier-discriminator and the dead-time free F1-TDC chip
characterized by high-resolution. The readout electronics has been
designed taking into account the high photon flux in the central part of
the detector and the requirement to run at high trigger rates of up to 100
kHz with negligible dead-time. The system is designed as a very compact
setup and is mounted directly behind the multi-anode photomultipliers.
The data are digitized on the frontend boards and transferred via optical
links to the readout system. The read-out electronics system is described
in detail together with its measured performances.
\vspace{-0.5cm}
\end{abstract}

\end{frontmatter}

\vspace*{-0.5cm}
% main text
\section{Introduction}
\label{Introduction}
\vspace*{-0.2cm}
The COMPASS experiment is a fixed target experiment at CERN. Its physics
program is focused on the investigation of the internal structure of the 
nucleon using muon and hadron beams. The main experimental  challenge is the
need to cope with high luminosity, resulting in high beam intensities and large
trigger rates. The setup of the COMPASS spectrometer is described in
\cite{Spectrometer}.

Charged hadron identification is obtained using a large Ring Imaging
Cherenkov detector, the COMPASS RICH-1 \cite{Rich-1}. The RICH detector uses
$C_4F_{10}$ as radiator gas inside a $5\times 6$~m$^2$ wide and $3$~meter deep
vessel. The produced Cherenkov photons are reflected on a $20$~m$^2$ mirror
wall onto two sets of photon detectors, an upper and a lower one. 

Until 2004, the photon detectors used were $8$ multi-wire proportional chambers
(MWPCs) with cesium iodide (CsI) photo-cathodes covering an active surface of
about $5.2$~m$^2$. Limitations to the RICH-1 performance at high rates
were related to the photon detector nature. Due to the presence of CsI
photocathodes, the MWPCs could not be operated at high gain, thus requiring a
long integration time of about $0.5$~$\mu$s of the front-end electronics, a
modified version of the GASSIPLEX chip. 

This limited the performance of the
photon detection in the central part of the detector  in two ways: In the
COMPASS experimental environment a large flux of halo muons 
accounts for about 10\% to 20\% of the total beam flux. At high beam intensities
of up to $10^8$ muons per second, the  halo muons create a considerable
background of Cherenkov photons. These photons create an uncorrelated
background on the photon detectors, which reduces the particle identification
efficiency and purity, especially for particles in the very forward direction.
Since also the base-line restoration of the GASSIPLEX output takes about
$3.5\,\mu$s, a large dead-time is created at high trigger rates. 

\section{RICH-1 Upgrade} 
\vspace*{-0.2cm} 
Therefore a new  and fast photon detection system was developed and installed
between Autumn 2004 and Spring 2006 in order to be able to distinguish by time
information between
photons from physics events and background,
and to be able to run at higher trigger rates of up to $100$~kHz. The upgrade of
the COMPASS RICH-1 is two-fold: In the central part of the photon detectors
(1/4 of the surface), the MWPCs have been replaced by $576$ Multi-Anode
Photo-Multipliers (MAPMTs) \cite{Fulvio} with new fast readout electronics,
which will be discussed in this contribution.  In the outer part, the existing
MWPCs have been equipped with a faster  readout electronics based on the APV
preamplifier with sampling ADCs \cite{APV}.

\section{Fast Readout Electronics for the MAPMTs} 
\begin{figure}
\includegraphics[width=0.5\textwidth]{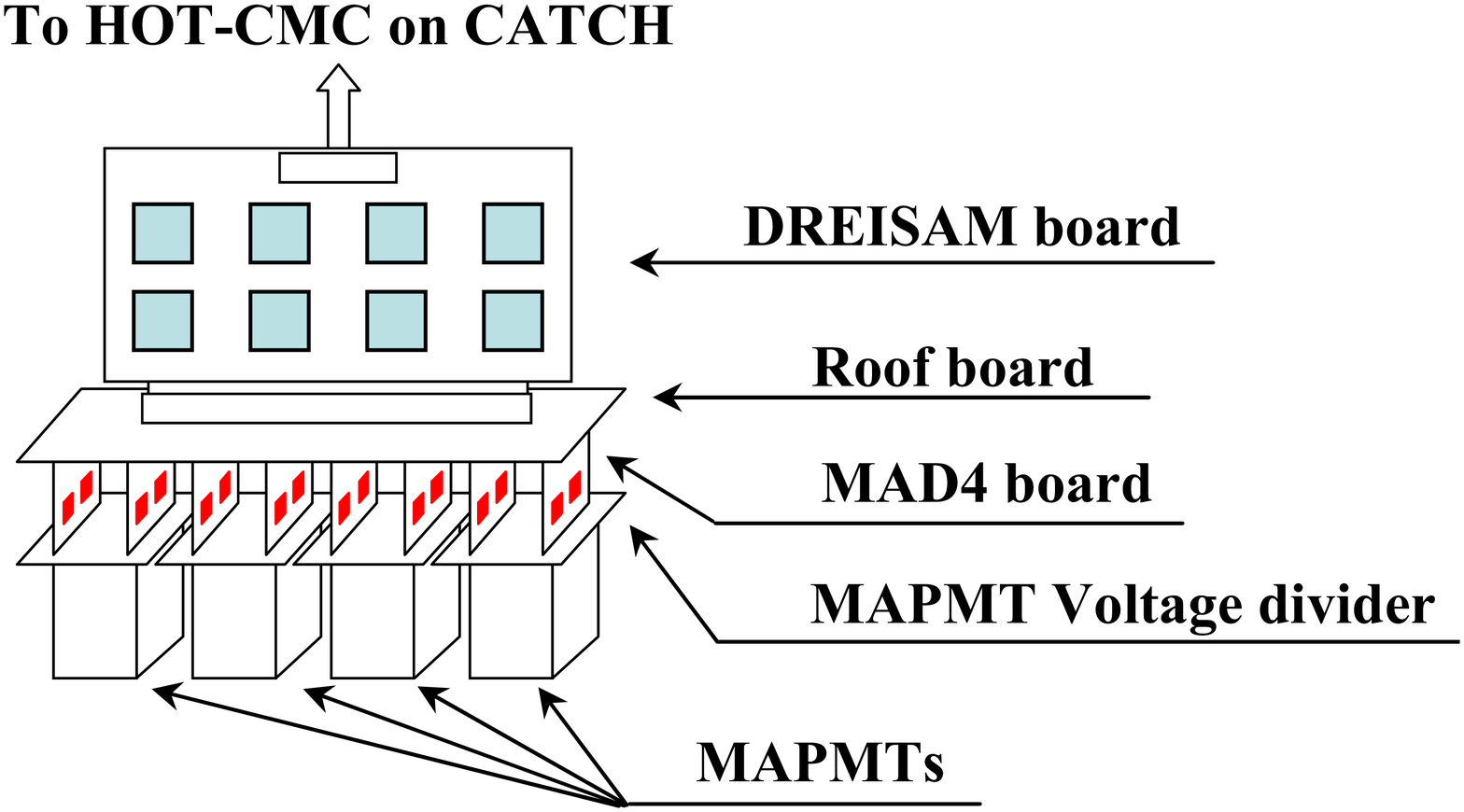}
\vspace*{-1.2cm}
\caption{Scheme of the read-out system.}
\label{fig1}
\end{figure}

\vspace*{-0.2cm} The MAPMTs used for a fast
photon detection in the central part of the photon detectors are 16-channel
multi-anode photomultipliers H7600-03-M16 from Hamamatsu \cite{Andy}. The
readout system \cite{MAD} for the MAPMTs is based on the MAD4
preamplifier-discriminator \cite{MAD2} and the dead-time free F1-TDC
characterized by a very good time resolution \cite{F1}. The electronics system is
mounted in a very compact setup as close as possible to the photomultipliers
(Fig.~\ref{fig1}). This minimizes the electrical noise and takes into account
the limited space in front of the RICH detector.

\section{Analogue Front-end Electronics} 
\begin{figure}
\includegraphics[width=0.5\textwidth]{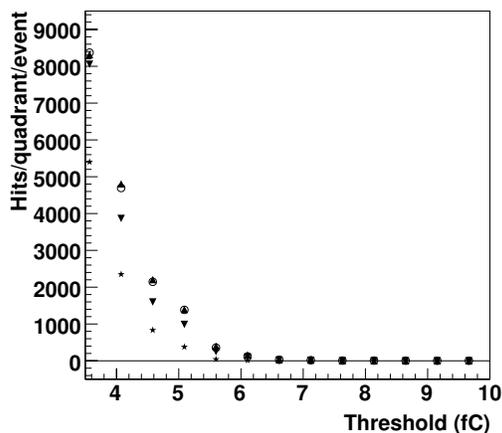}
\vspace*{-0.8cm}
\caption{Noise rate as a function of the threshold setting for one 
quadrant of the detector with $144$~MAPMTs.}
\label{threshold}
\end{figure}

\begin{figure}
\includegraphics[width=0.5\textwidth]{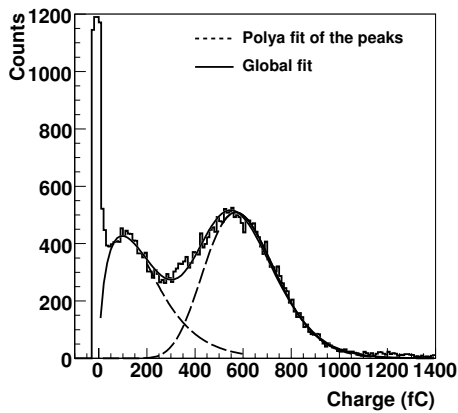}
\vspace*{-0.8cm}
\caption{Amplitude spectrum of the MAPMT for single photons at
$900$~V operating voltage, measured with an ADC.}
\label{signal}
\end{figure}

\begin{figure}
\includegraphics[clip, bb=0 0 800 600, width=0.5\textwidth]{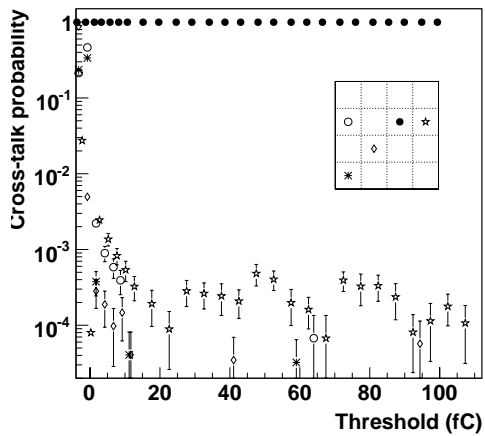}
\vspace*{-0.8cm}
\caption{Cross talk level as a function of the threshold setting, obtained by
illuminating in single photon mode a single pixel (black dot) with a focused 
laser beam.}
\label{fig2}
\end{figure}

\vspace*{-0.2cm} The analogue front-end board amplifies
the signal from the photomultiplier, discriminates it and sends it as a
differential signal to the digital board. Each front-end card is equipped with
$2$ MAD4 chips with $4$ channels each. The MAD4 chip features a
charge-sensitive preamplifier with fixed gain ($3.35$~mV/fC), a shaper and a
discriminator with digitally adjustable threshold. To match the amplitude of
the MAPMT signal to the input stage of the MAD4 chip, a resistive voltage
divider attenuates the signal by a factor of $2.4$. The binning of the threshold
setting 
was chosen to be $0.5$~fC/digit. The measured noise level is $<7$~fC
(Fig. \ref{threshold}), while typical MAPMT signals have amplitudes between
$100$ and  $1000$~fC (Fig. \ref{signal}) \cite{Andy}. The signal peak at lower
amplitudes originates from photoelectrons which are
missing one amplification stage in the MAPMT. Since the signal fraction 
of this peak is significant, a threshold setting below this peak 
is essential for achieving high efficiency.  A typical
threshold setting of about $40$~fC is chosen. For this threshold, the
excellent signal-to-noise ratio allows to obtain a very high efficiency,
preserving a negligible level of cross-talk (Fig. \ref{fig2}). 

The MAD4 chip can operate at input rates up to $1$~MHz per
channel. An upgraded version of the MAD4 chip is under development, 
capable of input rates up to $5$~MHz, the CMAD.


\section{Digital Front-end Electronics}  
\begin{figure}[b]
\includegraphics[width=0.5\textwidth]{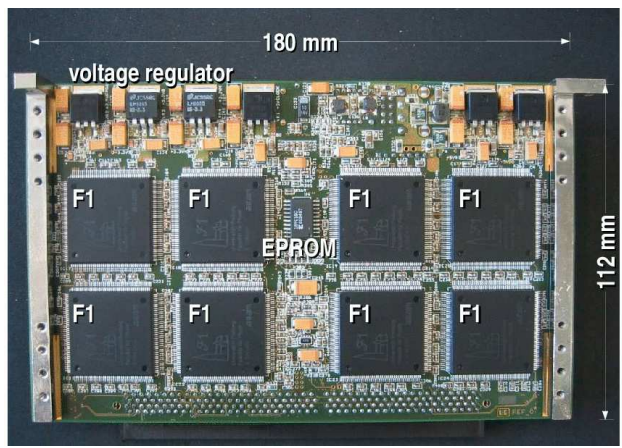}
\vspace*{-0.6cm}
\caption{Front side of the DREISAM card with 8 F1-TDC chips.}
\label{Dreisam}
\end{figure}

\vspace*{-0.2cm} The digital part of the new RICH-1
frontend electronics consists of the DREISAM front-end board, which is equipped
with eight F1-TDC chips to read out four MAPMTs (Fig. \ref{Dreisam}). The
board was designed in a very compact way. The data are
digitized on the DREISAM board and are sent out via optical links to the
HOT-CMC board, a small mezzanine card on the CATCH, the common readout-driver
board of the COMPASS experiment. The F1 chips on the digital boards have a
digitization bin-width of $108.3$~ps
and can operate at input data rates of up to $10$~MHz per channel. The readout of
the data can be performed at trigger rates up to $100$~kHz. 

\section{System Performances}
\vspace*{-0.2cm}
The time resolution of the complete system consisting of MAPMT, MAD4 board and
DREISAM board has been determined by illuminating the MAPMT photocathode with 
optical pulses of width less
than $50$~ps from a pulsed laser system. The laser intensity was attenuated by optical
filters to obtain single photon signals on the MAPMT. The time resolution of
the complete system was determined to be  $\sigma=320$~ps.
 
\begin{figure}
\includegraphics[width=0.5\textwidth]{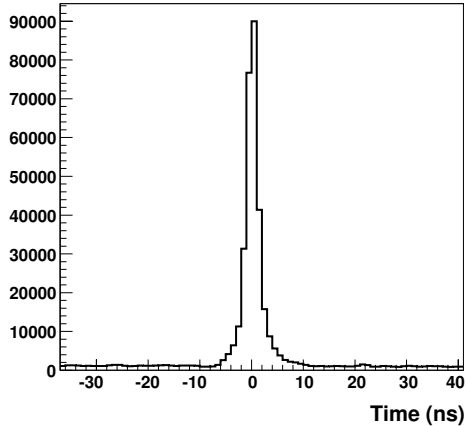}
\vspace*{-0.8cm}
\caption{Physics signal and background from 2006  data.}
\label{time}
\end{figure}

\begin{figure}
\includegraphics[width=0.5\textwidth]{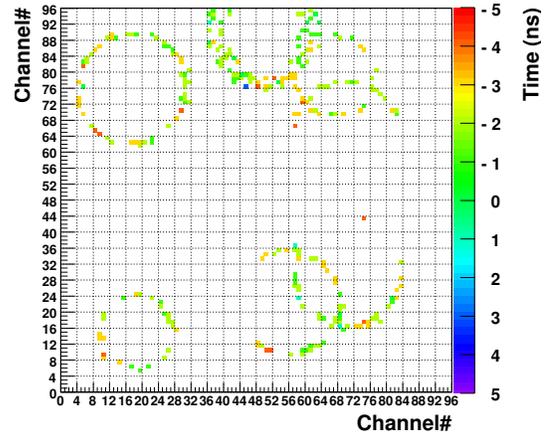}
\vspace*{-0.8cm}
\caption{Online display of Cherenkov rings from a physics event.}
\label{Rings}
\end{figure}

The upgraded photon detection system of the COMPASS RICH-1 has been stably 
operated during the beam-time in 2006 and 2007. In Fig. \ref{time} the time
spectrum of the detected Cherenkov photons is shown. The central peak of the physics
signal has a standard deviation of about $1$~ns. The background below the peak is created by
uncorrelated Cherenkov photons mainly from muon-beam halo particles.  
The observed width of the central peak is determined by the different
geometrical path length of the photons in a Cherenkov ring traveling from 
the mirrors to the photon detection system. This has been confirmed by a Monte
Carlo simulation of the detector setup. 
By applying a suitable off\-line time-cut of $\pm 5$~ns around the signal peak,
an excellent background suppression is achieved. Cherenkov rings from a
physics event are shown in Fig. \ref{Rings}.

\section{Conclusions} 
A fast front-end electronics for the read-out of the
MAPMTs of the COMPASS RICH-1 was designed and successfully installed in 2006.
The new electronics has an excellent time resolution for Cherenkov photons  of
less than $1$~ns and thus allows a very good background suppression of 
 photons originating from halo  muons. Its high hit capability allows a dead-time free
data taking at trigger rates of up to $100$~kHz.  The upgraded detector and the new
electronics entirely fulfill the expected performances and have been operated successfully since the data taking period in 2006. With the upgraded
detector, the number of detected Cherenkov photons for saturated rings 
has increased from $14$
before to $56$ after the upgrade. The resolution of the
Cherenkov angle has improved from $0.6$ to $0.3$~mrad \cite{Federica}.
\vspace*{-0.8cm}

\section*{Acknowledgments}
We acknowledge the support from CERN and the support by the BMBF (Germany) and
the European Community-research Infrastructure Activity under the FP6 program
(Hadron Physics).

\end{document}